\documentclass[aps,amssymb,pra,showpacs,twocolumn,floatfix]{revtex4}

\usepackage{epsf}
\usepackage{amsmath}
\usepackage{graphicx}
\begin{document}
         
\title {Nuclear threads}

\author{B. Ivlev}

\affiliation
{Instituto de F\'{\i}sica, Universidad Aut\'onoma de San Luis Potos\'{\i},
San Luis Potos\'{\i}, 78000 Mexico\\}

\begin{abstract}

The internal structure of the deuteron weakly influences a motion of its center of mass (translational motion). The scenario can be different 
when the translational wave function has a formal singularity along the line (thread) connecting the deuteron and another Coulomb center. The 
singularity is cut off by fluctuations of nuclear internal degrees of freedom and the wave function becomes smooth with the narrow peak. This 
peak is associated with the certain potential well which is localized on the thread and dominates the Coulomb barrier. As a result, the deuteron 
is able to move along the thread with no classical reflections. Two deuterons can get in contact despite their energy is low. This is not an 
underbarrier motion but one along the region which became classically allowed due to the thread well. In experiments nuclear threads can be 
formed by the incident particle flux if to adjust its angular distribution at large distances.

\end{abstract} \vskip 1.0cm


\maketitle

\section{INTRODUCTION}
\label{intr}
The scattering problem in quantum mechanics is a study of a particle flux coming from large distances 
where it is related to an almost plane wave \cite{LANDAU1,NEWTON}. A motion of two particles can be 
reduced to scattering in the frame of center of mass. This is possible when each particle is a point 
charge. In this case one can apply the usual Schr\"{o}dinger equation accounting for a potential 
interaction of two particles. Results are well described in textbooks \cite{LANDAU1,NEWTON}.

Suppose the vector $\vec R$ (translational coordinate) to connect the centers of mass of two deuterons. When the distance between two deuterons 
$\vec R$ is much larger than the nuclear radius $r_N$ one can use an approximation of point charges interacting via their Coulomb field. The 
wave function $\Psi^{(0)}_{C}(\vec R)\psi_{N}$ is a product of the usual Coulomb part \cite{LANDAU1} and the nuclear one, $\psi_N$, for 
non-interacting nuclei. Corrections to this result are small.

Those statements look obvious since $\Psi^{(0)}_{C}(\vec R)$ is smooth and hardly varies within the deuteron size. However that scenario may be 
not correct when the translational wave function varies rapidly. This can happen when that function has a singularity on some line. At the first 
sight, such wave function is not physical. But the situation is more complicated.

There are various cases when the wave function is singular on some line. The simplest example is connected 
with complex angular momenta. As known, a wave function, defined for discrete angular momenta $l=0,1,2,...$, 
can be analytically continued in the complex $l$ plane \cite{REGGE,LANDAU1}. But instead, one can take a 
solution  with only one $l$ which is chosen complex. The related state is localized within some interval of 
polar angles $\theta$ with respect to the $z$ axis. Sec.~\ref{form}. This formal solution of the Schr\"{o}dinger equation is 
not physical since it has the logarithmic (for the axially symmetric case) singularity on the $z$ axis due 
to non-integer $l$. 

The logarithmic singularity of the wave function is a consequence of the quantum mechanical description in terms of particle coordinate 
$\vec R$. But the deuteron has the finite radius $r_N\simeq 2.13\times 10^{-13}\,{\rm cm}$ \cite{POHL} and the internal structure 
\cite{CASTEN,WIRINGA,SLAUS,ORD,JACK,COTT,NAG,MACH,FOREST}. Due to the Coulomb interaction, a motion of the deuteron center of mass is not
separated from its internal degrees of freedom. The latter provide spatial fluctuations in the singular state resulting in its smearing.
This is analogous to electron ``oscillations'' in external fields, mediated by photons, in quantum electrodynamics \cite{LANDAU3}. In the both 
cases the particle momentum is not conserved in virtual processes. 

That mechanism results in a distribution of singularity lines within the nuclear radius in the vicinity of the $z$ axis (thread). Sec.~\ref{cone}. 
The wave function of the deuteron translational motion becomes smooth and having a peak on the thread. Such peak is associated with the certain 
potential well which is localized on the thread. The well is formed jointly by the coordinate $\vec R$ and internal nuclear subsystem. The 
resulting potential is a combination of Coulomb and thread ones. The Coulomb barrier is dominated by the thread well and the motion toward the 
Coulomb center occurs with no classical reflections. 

In the motion along the thread two deuterons can get in contact despite they are of low energy. This is not an underbarrier motion but one along 
the region which became classically allowed due to the thread well. Nuclear thread can be created at some region by adjusting a deuteron flux at 
large distances. Sec.~\ref{exp}. 

The form of that flux plays a role of driving force for the phenomenon since it provides the large increase of the wave function at a short 
range close to the thread. At these distances internal nuclear subsystem enters the game. So in this manner the bridge is built up 
between macroscopic and internal nuclear subsystems. This is unusual since the former are related to room temperature energy but the latter are 
associated with high energy of the nuclear scale. 

Actually details of the wave function and the well close to the thread are not crucial and serve for interpretation of the mechanism. Namely how
the logarithmic singularity is cut off. The particle flux to the center is mainly determined by the outer region of the thread described by the
usual quantum mechanics for the translational motion. In that region the wave function increases toward the thread which is equivalent to the 
negative transverse kinetic energy. This energy part dominates the Coulomb barrier. 

A continuous superposition of radial waves, propagating to the same point but with different directions 
in space, is also a state. It can be extended over a macroscopic space domain, say, of micron size where it is
smooth (macroscopic nuclear state). The wave function is mainly the product $\Psi_C(\vec R)\psi_N$ originated from a spatial average of 
the regions outside threads. The difference of that form from one, calculate for point charges \cite{LANDAU1}, is the 
Coulomb part. $\Psi_C(\vec R)$ strongly differs from $\Psi^{(0)}_{C}(\vec R)$. 

That wave function does not obey the usual Schr\"{o}dinger equation in terms of translational coordinates. But there is no contradiction 
since the wave function is determined by the total system accounting for also internal nuclear degrees of freedom. There is an additional 
contribution to the wave function where the translational motion and internal nuclear processes are not separated. This part originates from 
a spatial average of inner regions of threads. That contribution is small (as the nuclear radius) but is connected to the high energy nuclear 
system. As a result, it substantially corrects the solution $\Psi^{(0)}_{C}(\vec R)$ of the Schr\"{o}dinger equation.

So a motion of nuclei in a macroscopic region of space becomes completely different compared to the usual 
expectation based on the point charge approach. (Sec.~\ref{super}).
\section{COMPLEX ANGULAR MOMENTA}
\label{form}
In this section we consider a singular wave function related to complex angular momentum. A motion of two deuterons in the frame of their center 
of mass is described by the Schr\"{o}dinger equation
\begin{equation}
\label{1}
-\frac{\hbar^2}{M_D}\,\frac{\partial^2\psi_C}{\partial\vec R^2}+\frac{e^2}{R}\,\psi_C=E\psi_C,
\end{equation}
where the deuteron mass $M_D$ is double of the reduced mass. The index of the wave function is 
referred to the Coulomb system described by the translation $\vec R$ of point particles. Note that the nuclear Compton length 
$\hbar/cM_D\simeq 1.05\times 10^{-14}\,{\rm cm}$ is less than the deuteron radius. So non-relativistic quantum mechanics is applicable.

It is convenient to introduce the large semiclassical 
parameter 
\begin{equation}
\label{2}
B=\frac{e^2}{\hbar}\sqrt{\frac{M_D}{E}}
\end{equation}
and to measure distance in the units of $2e^2/E$. The deuteron mass is given by the relation $M_Dc^2\simeq 1.87\times 10^3\,{\rm MeV}$. At low 
energy $E=300~{\rm Kelvin}\simeq 2.58\times 10^{-2}\,{\rm eV}$ one can estimate $B\simeq 1965$ and $2e^2/E\simeq 1114~\AA$. 

Below we consider the axially symmetric wave function which satisfies the equation in the dimensionless 
form
\begin{eqnarray}
\nonumber
&&-\frac{1}{4B^2R^2}\left[\frac{\partial}{\partial R}\left(R^2\frac{\partial\psi_C}{\partial R}\right)
+\frac{1}{\sin\theta}\,
\frac{\partial}{\partial\theta}\left(\sin\theta\frac{\partial\psi_C}{\partial\theta}\right)\right]\\
\label{3}
&&+\frac{\psi_C}{2R}=\psi_C,
\end{eqnarray}
where $R^2=x^2+y^2+z^2$ and $\theta$ is the angle between $\vec R$ and the $z$ axis (polar angle). A 
solution of Eq.~(\ref{3}) can be written as \cite{LANDAU1}
\begin{equation}
\label{4}
\psi_C(R,\theta)=\frac{\Phi_l(R)}{R}\,P_l(\cos\theta)\,,
\end{equation}
where $P_l(\cos\theta)$ satisfies the same equation as the Legendre polynomial
\begin{equation}
\label{5}
\frac{1}{\sin\theta}\,
\frac{\partial}{\partial\theta}\left(\sin\theta\frac{\partial P_l}{\partial\theta}\right)+l(1+l)P_l=0\,.
\end{equation}
The solution of Eq.~(\ref{5}) is \cite{LANDAU1}
\begin{equation}
\label{6}
P_l(\cos\theta)=\exp\left[-i\pi\left(l+\frac{1}{2}\right)\right]
F\left(-l,1+l,1,\frac{1+\cos\theta}{2}\right)
\end{equation}
where
\begin{equation}
\label{7}
F(\alpha,\beta,\gamma,z)=1+\frac{\alpha\beta}{\gamma}\frac{z}{1!}
+\frac{\alpha(\alpha+1)\beta(\beta+1)}{\gamma(\gamma+1)}\frac{z}{2!}+...
\end{equation}
is the hypergeometric function. The function $\Phi_l(R)$ obeys the equation
\begin{equation}
\label{8}
-\frac{1}{4B^2}\frac{\partial^2\Phi_l}{\partial R^2}+\left[\frac{1}{2R}
+\frac{l(1+l)}{4B^2R^2}\right]\Phi_l=\Phi_l\,.
\end{equation}
As known, the solution of Eq.~(\ref{8}) is
\begin{eqnarray}
\nonumber
&&\Phi_l(R)=a_lR^{1+l}\exp(-2iBR)\\
\label{9}
&&\times F\left(1+l-\frac{iB}{2},\,2+2l,\,4iBR\right),
\end{eqnarray}
where
\begin{equation}
\label{10}
F(\alpha,\gamma,z)=1+\frac{\alpha}{\gamma}\frac{z}{1!}
+\frac{\alpha(\alpha+1)}{\gamma(\gamma+1)}\frac{z}{2!}+...
\end{equation}
is the confluent hypergeometric function \cite{LANDAU1} and we choose
\begin{eqnarray}
\nonumber
a_l=-\frac{iB}{l}\sqrt{2
+\frac{iB}{l}}\exp\bigg[\left(l-\frac{iB}{2}\right)\left(1-\frac{i\pi}{2}\right)\\
\label{11}
+\frac{iB}{2}\ln\frac{2l+iB}{8B}+l\,\ln\frac{B(2l+iB)}{2l^2}\bigg].
\end{eqnarray}
\subsection{Angular part $P_l(\cos\theta)$}
The polar angle $\theta$ is in the interval $0<\theta<\pi$. At $\theta=\pi$ the angular part is 
$P_l=\exp[-i\pi(l+1/2)]$. At $(\pi-\theta)\ll 1$ Eq.~(\ref{5}) turns into the Bessel equation with the 
solution $P_l(\cos\theta)\sim J_0[(\pi-\theta)\sqrt{l(1+l)}]$. We consider the limit 
$1\ll l$. At $1/l\ll(\pi-\theta)\ll 1$, according to the asymptotic of the Bessel function, the solution 
has the form
\begin{eqnarray}
\nonumber
P_l(\cos\theta)=\sqrt{\frac{2}{\pi l\sin\theta}}\,\exp\left[-i\pi\left(l+\frac{1}{2}\right)\right]\\
\label{12}
\times\cos\left[(\pi-\theta)\sqrt{\left(l+\frac{1}{2}\right)^2-\frac{1}{4}}-\frac{\pi}{4}\right].
\end{eqnarray}
The equation (\ref{12}) holds, more generally, at $1/l\ll\sin\theta$. To obtain a solution at 
$\theta\ll 1/l$, one can use the expression of the 
hypergeometric function when its argument is close to unity. It reads at $(1-\xi)\ll 1/l^2$ \cite{LANDAU1}
\begin{equation}
\label{13}
F(-l,1+l,1,\xi)=\frac{\sin\pi l}{\pi}\ln\Bigg\{\left[\frac{1}{4}
-\left(l+\frac{1}{2}\right)^2\right](1-\xi)\Bigg\}
\end{equation}
For usual Legendre polynomials $l$ is an integer number and the logarithmically divergent term (\ref{13}) 
is absent. Below we consider 
\begin{equation}
\label{14}
l=-\frac{1}{2}-iBq\,,
\end{equation}
where the real $q$ satisfies the condition $1<2q$. The parameter $q$ is a characteristic number of the problem. 
The above Bessel function goes over into the function 
$I_0(BQ\theta)$ \cite {GRAD}. This allows to write the asymptotics
of the angular part in the forms
\begin{equation}
\label{15}
P_l(\cos\theta)\simeq\frac{1}{\sqrt{2\pi Bq\theta}}\exp(-Bq\theta),
\hspace{0.4cm}\frac{1}{B}\ll\theta\ll 1
\end{equation}
\begin{equation}
\label{16}
P_l(\cos\theta)\simeq\frac{1}{\pi}\ln\frac{2}{Bq\theta}\,,\hspace{0.4cm}\theta\ll\frac{1}{B}\,.
\end{equation}
The function $P_l(\cos\theta)$ under the condition (\ref{14}) is real and positive.
\subsection{Radial part $\Phi_l(R)$}
The semiclassical solution of Eq.~(\ref{8}) holds at $1\ll RB$ and has the form \cite{LANDAU1}
\begin{eqnarray}
\nonumber
&&\Phi_l(R)=b(q)\left(1-\frac{1}{2R}+\frac{q^2}{4R^2}\right)^{-1/4}\\
\nonumber
&&\times\Bigg[\exp\left(-2iB\int^{R}_{1/4}dR_1\sqrt{1-\frac{1}{2R_1}+\frac{q^2}{4R^{2}_{1}}}\,\right)+\\
\label{17}
&&s(q)\exp\left(2iB\int^{R}_{1/4}dR_1\sqrt{1-\frac{1}{2R_1}+\frac{q^2}{4R^{2}_{1}}}\,\right)\Bigg],
\end{eqnarray}
where $b(q)$ and $s(q)$ are the certain functions to be determined. The wave function (\ref{17}) contains 
the incident wave (the first term) and the reflected wave (the second term). The classical turning points 
$R_t$ are defined by the condition $1-1/2R+q^2/4R^2=0$ and are given by the relation
\begin{equation}
\label{18}
R_{t1,2}=\frac{1}{4}\left(1\pm i\sqrt{4q^2-1}\right).
\end{equation}
Since $1<2q$, the both turning points are complex. The reflected wave is generic with one in overbarrier 
reflection. It is generated on the Stokes line \cite{HEAD} which connects two turning points (\ref{18}). 
The amplitude of the reflected wave, $s(q)$, is exponentially small in the semiclassical limit $1\ll B$ 
\begin{equation}
\label{19}
s(q)\sim \exp\left[-\pi B\left(q-\frac{1}{2}\right)\right].
\end{equation}
The result (\ref{19}) can be obtained directly from asymptotics of the confluent hypergeometric function in
Eq.~(\ref{9}). Another way is to use the integration contour in the complex $R$ plane bypassing the 
turning point from the real axis \cite{LANDAU1,HEAD}.
\subsection{Semiclassical wave function}
The only contribution to the wave function comes from the angular momentum (\ref{14}) where $q>1/2$. We 
suppose $q\sim 1$. Neglecting the small reflected wave, one can write the wave function in the form
\begin{equation}
\label{20}
\psi_C(R,\theta)=\frac{\Phi_l(R)}{R}\,P_l(\cos\theta).
\end{equation}

In the semiclassical approximation, $1/B\ll R$, this wave function is \cite{LANDAU1} 
\begin{eqnarray}
\nonumber
\psi_C(R,\theta)=\frac{b(q)}{R}\,P_l(\cos\theta)\left(1-\frac{1}{2R}+\frac{q^2}{4R^2}\right)^{-1/4}\\
\label{21}
\times\exp\left(-2iB\int^{R}_{1/4}dR_1\sqrt{1-\frac{1}{2R_1}+\frac{q^2}{4R^{2}_{1}}}\,\right).
\end{eqnarray}
In order to match Eq.~(\ref{21}) and Eq.~(\ref{9}) with the definition (\ref{11}) one has to choose the 
form
\begin{eqnarray}
\nonumber
b(q)=\exp\Bigg\{-\frac{iB}{2}\Bigg[1+\sqrt{4q^2-1}+\ln\frac{8}{\sqrt{4q^2-1}}\\
\label{22}
+2q\ln\left(\frac{1}{2q+\sqrt{4q^2-1}}\sqrt{\frac{2q-1}{2q+1}}\right)\Bigg]\Bigg\}.
\end{eqnarray}
The form (\ref{22}) of $b(q)$ results in the asymptotics
\begin{equation}
\label{23}
\psi_C(R,\theta)=\frac{P_l(\cos\theta)}{R}\exp\left(-2iBR+\frac{iB}{2}\ln R\right),\hspace{0.1cm}1\ll R
\end{equation}
\begin{equation}
\label{24}
\psi_C(R,\theta)=P_l(\cos\theta)\sqrt{\frac{2}{qR}}\exp\left(-iBq\ln R+iB\chi\right),\hspace{0.1cm}R\ll 1
\end{equation}
\begin{figure}
\includegraphics[width=8.5cm]{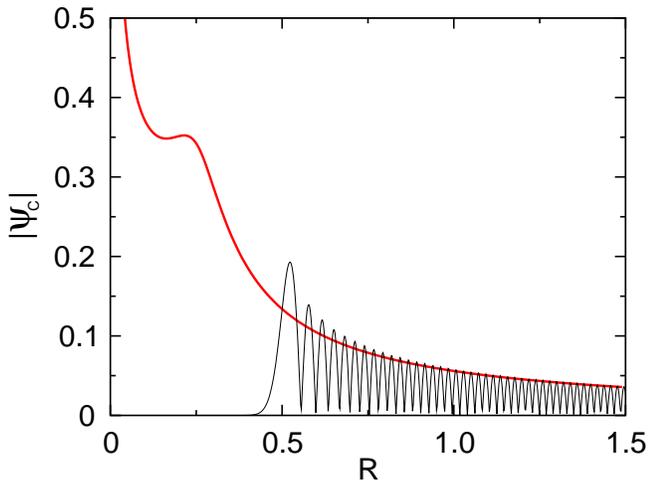}
\caption{\label{fig1} The Coulomb wave function is numerically plotted for $B=1000$, $q=0.526$, and 
$B\theta=0.1$. The usual isotropic WKB wave function (the oscillatory curve) is drawn for comparison.}
\end{figure}
where
\begin{equation}
\label{25}
\chi=\frac{1}{2}\left(\ln\frac{2q-1}{8}+2q\ln\frac{2q^2}{2q-1}-1-2q\right).
\end{equation}
The radial part of the wave function at large distances (\ref{23}) differs from the free radial wave by 
the logarithmic Coulomb phase \cite{LANDAU1}. At small $R$ one estimates $|\psi_C|\sim 1/\sqrt{R}$ which
does not contradict to the normalization condition. Such singularities are possible. An example is a 
solution of the Dirac equation.

The Coulomb wave function $\psi_C$ is numerically calculated on the basis of Eqs.~(\ref{6}) and (\ref{9}). 
This function is plotted in Fig.~\ref{fig1} and it well coincides with the semiclassical form (\ref{21}). 
$\psi_C$ in Fig.~\ref{fig1} does not decay under the Coulomb barrier. The formal reason for that is the negative
centrifugal energy which dominates the Coulomb barrier.

\begin{figure}
\includegraphics[width=8.5cm]{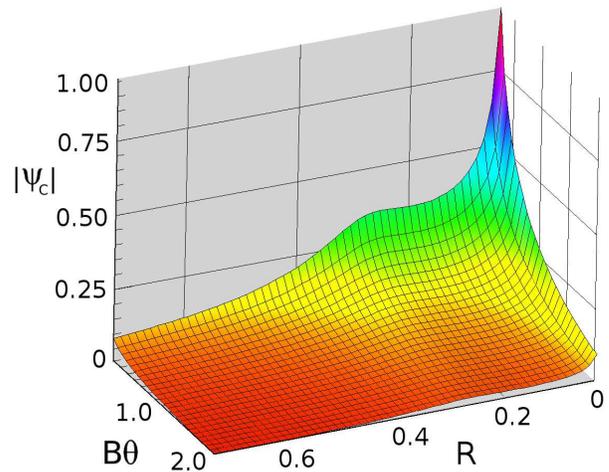}
\caption{\label{fig2} Three-dimensional plot of $|\psi_C|$. The wave function is localized within the
narrow cone of the angle $\theta\sim 1/B$.}
\end{figure}

The three-dimensional plot of $\psi_C$ is shown in Fig.~\ref{fig2}. The wave function is localized within 
the cone of the angle $\theta\sim 1/B$ according to Eqs.~(\ref{15}) and (\ref{16}).
\subsection{WKB approach}
Unlike the above forms, an incident flux of particles at large distances can be of the usual type, namely, related to 
the certain discrete quantum numbers. For example, it can be a plane wave or an isotropic one. See also 
\cite{COLEMAN1,COLEMAN2,SAKITA,LEGGETT,HELLER,ANKER,SCHMID1,SCHMID2}. In this case, if an energy is not high, the 
probability of tunneling through the Coulomb barrier $w_{WKB}\sim\exp(-2\pi\lambda\sqrt{2}/r_{B})$ is extremely 
small according to the theory of Wentzel, Kramers, and Brillouin (WKB) \cite{LANDAU1}. Here 
$\lambda=\hbar/\sqrt{2M_{D}E}$ is the de Broglie wave length, 
$r_{B}=2\hbar^{2}/M_{D}e^{2}\simeq 2.9\times 10^{-12}~{\rm cm}$ is the nuclear Bohr radius, and $E$ is the energy of 
two deuterons. The energy dependence of the WKB tunneling probability is clear from the relation
\begin{equation}
\label{25a}
w_{WKB}\sim\exp\left(-\frac{\pi e^2}{\hbar}\sqrt{\frac{M_D}{E}}\,\right).
\end{equation}
When $E$ corresponds to the room temperature of $300~{\rm Kelvin}$, the wave length is $\lambda\simeq 0.20~\AA$ and the tunneling probability 
(\ref{25a}) is $10^{-2681}$ according to usual estimates.

In Fig.~\ref{fig1} the isotropic WKB wave function is drawn for comparison. It exponentially decays under the barrier. 
Due to interference of the incident and reflected waves the modulus of the WKB function oscillates. To resolve the 
oscillations the moderate value, $B=50$, in Fig.~\ref{fig1} is chosen for the WKB wave function.
\subsection{Particle flux}
The particle flux in the $\theta$ direction is zero since $P_l(\cos\theta)$ is real. The radial particle 
flux is 
\begin{equation}
\label{26}
I_R=2\pi R^2\int^{\pi}_{0}d\theta\sin\theta\,{\rm Im}\left(\psi^{*}\frac{\partial\psi}{\partial R}\right).
\end{equation}
A typical angles are small, $\theta\sim 1/B$ allowing to extend the integration from zero to infinity. 
Therefore, using Eqs.~(\ref{23}) and (\ref{24}), one can write the particle flux on the center as
\begin{equation}
\label{27}
I_R=-\frac{4\pi}{Bq^2}\int^{\infty}_{0}P^{2}_{l}(\cos\theta)(Bq\theta)d(Bq\theta)=-\frac{c_1}{Bq^2}\,,
\end{equation}
where the constant $c_1$ is positive and of the order of unity. The radial current density is proportional 
to $1/R^2$ and the total radial current $I_R$ does not depend on $R$. Due to particle conservation, there 
is a counter-flow. Its amplitude depends on a fraction of deuterons getting out of the game due to possible nuclear reactions.  

The total incident flux propagates within the narrow cone of the angle $\theta\sim 1/B$. The wave function
(\ref{20}) can be represented in the form 
\begin{equation}
\label{28}
\psi_C(R,\vec n)=\frac{\Phi_l(R)}{R}\,P_l(\vec n\vec n_z)\,,
\end{equation}
where $\vec R=R\vec n$ and $\vec n$ is the unitary vector in the space, related to the polar angle 
$\theta$ and the azimuthal angle $\varphi$. The unitary vector $\vec n_z$ is directed along the $z$ axis 
making the angle $\theta$ with the direction $\vec n$. Eqs.~(\ref{28}) and (\ref{20}) are valid at $r_N<R\theta$. 
\section{THREAD STATE OF NUCLEI}
\label{cone}
\begin{figure}
\includegraphics[width=6cm]{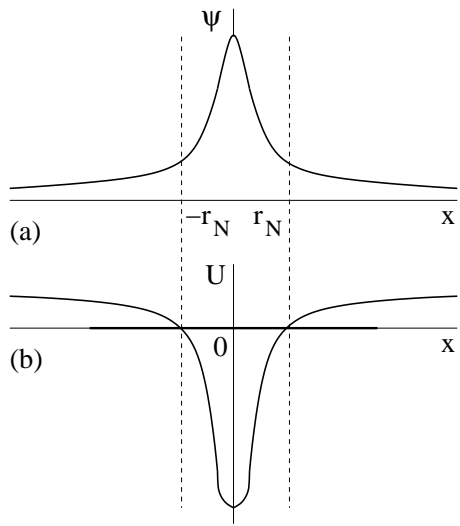}
\caption{\label{fig3} Features of the thread state at $y=0$. Dashed lines restrict the thread region. (a) Thread wave function. (b) Well 
associated with the thread. The effective energy level (see Eq.~(\ref{28a})) is shown by the thick line.}
\end{figure}
Usually the wave function, related to the translational motion, is smooth at distances of the nuclear size. In this case translational  
and nuclear internal subsystems are separated. The situation becomes different when the translational wave function varies rapidly on 
the nuclear scale. This is the case of the wave function $\psi\sim\ln(1/r)$ of Sec.~\ref{form} where $r^2=x^2+y^2$. It strongly increases 
approaching the $z$ axis which is the singularity line. But even in this case, not too close to the $z$ axis $r_N<r$, the above subsystems are 
still separated and the wave function has the form
\begin{equation}
\label{28aaa}
\psi=\psi_C(\vec R)\psi_{N}.
\end{equation}
Here $\psi_C(\vec R)$ with the eigenvalue $E$ is calculated in Sec.~\ref{form} and $\psi_{N}$ corresponds to the equilibrium nuclear state with 
the deuteron energy $M_Dc^2\simeq 1.87\times 10^3~{\rm MeV}$. This energy is smaller than one of bare proton plus neutron by the binding energy 
$E_b\simeq 2.225~{\rm MeV}$. The binding energy is associated with the equilibrium charge-mass distribution inside the deuteron  
\cite{CASTEN,WIRINGA,SLAUS,ORD,JACK,COTT,NAG,MACH,FOREST}. 

At small $\theta=r/R\ll 1$ the equation (\ref{5}), with (\ref{14}), can be written in the form
\begin{equation}
\label{28a}
-\frac{\hbar^2}{M_Dr}\frac{\partial}{\partial r}\left(r\frac{\partial\psi}{\partial r}\right)+U\psi=0,
\end{equation}
\begin{figure}
\includegraphics[width=8cm]{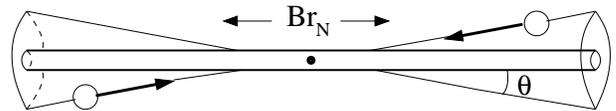}
\caption{\label{fig4} Nuclear thread of the nuclear radius along the $z$ axis connects two deuterons. The hybridization of translational and 
internal nuclear subsystems occurs inside the thread. Radial beams of deuterons are pointed at the dot.}
\end{figure}
which holds at $r_N<r$. In this region $U=q^2e^4/(ER^2)$ does not depend on nuclear variables. The wave function (\ref{28aaa}) satisfies 
Eq.~(\ref{28a}) where $R$ is fixed. According to (\ref{16}), the solution of Eq.~(\ref{28a}) has the logarithmic singularity at $r=0$.
\subsection{Thread wave function}
Let us formally consider the situation when the entire deuteron charge would be localized at the certain point $\vec R+\vec u$. Then one can 
shift the system by $\vec u$ to get $\vec R+\vec u$ as a new radial vector and the singularity line is also shifted by the vector $\vec u$. The 
wave function $(-1/2\pi)\ln|\vec r+\vec u_{\perp}|$ results in the term $(\hbar^2/M_D)\delta(\vec r+\vec u_{\perp})$ in the Hamiltonian 
(\ref{28a}). The vector $\vec u_{\perp}$ does not have the $z$ component. 

The deuteron charge, localized at two points $\vec R+\vec u_1$ and $\vec R+\vec u_2$, leads to two logarithmic singularities 
$\ln|\vec r+\vec u_{1,2\perp}|$. The singularity line cannot have another position since it should be terminated at a Coulomb center. The 
continuous charge distribution inside the deuteron results in a continuous distribution of singularity lines. This is equivalent to the smooth 
wave function as in Fig.~\ref{fig3}(a).

The charge density inside the deuteron is a matter of fluctuations with the typical time $\hbar/E_b$. Close to the Coulomb center, when
$r_N\lesssim R$, the Coulomb energy $e^2/r_N\simeq 0.68~{\rm MeV}$ is of the order of $E_b$. There is the Coulomb interaction 
$U_{CN}=\int d^3ue\rho/|\vec R+\vec u|$, $u\sim r_N$, depending on internal nuclear coordinates through the nuclear charge density $\rho$. This 
interaction results in non-separation of translation and internal nuclear coordinates. The latter provide spatial fluctuations in the singular 
state resulting in its smearing as noted above.

This is analogous to the electron dynamics in an external field, mediated by photons, in quantum electrodynamics \cite{LANDAU3}. See 
Appendix. In the both cases the particle momentum is not conserved in virtual processes leading to spatial ``oscillations'' of the center of mass.

The state is no more characterized by only one singularity line. Since these lines are continuously distributed within the deuteron size the 
wave function can be written as
\begin{equation}
\label{28A}
\psi=\psi_{N}\int\frac{d^2u_{\perp}}{r^{2}_{N}}\alpha(\vec u_{\perp})\psi_C(\vec R+\vec u_{\perp})+\delta\psi.
\end{equation}
The function $\alpha(\vec u_{\perp})$ is localized at $u_{\perp}\lesssim r_N$. It is easy to see that smearing (\ref{28A}) leads, instead of 
the $\delta$-function, to the term 
\begin{equation}
\label{28BB}
\frac{2\hbar^2}{M_Dr^{2}_{N}}\frac{\Phi_l(z)}{z}\alpha(-\vec r)\psi_{N}
\end{equation}
in the left hand side of Eq.~(\ref{28a}). This means that $\delta\psi$ is a reaction of the nuclear system on the perturbation (\ref{28BB}). If 
the nuclear system would be described by a Schr\"{o}dinger equation, $\delta\psi$ is determined by the integration with Green's function. 
$\delta\psi$ is localized on the distance $r_N$ close to the $z$ axis. 

At not small distances from the thread, fluctuation corrections to the deuteron wave function (\ref{28aaa}) are small. This function is kept by 
the external conditions and cannot have a singularity at smaller distances. Otherwise the singularity would be smeared out by the uncertainty in
positions of the deuteron center of mass.

At some typical values of nuclear internal degrees of freedom the resulting wave function is schematically plotted in Fig.~\ref{fig3}(a). At 
$r_N\ll r,z$ the wave function contains mainly $\ln r$ plus small corrections due to quantum fluctuations of internal nuclear degrees of freedom. 
At $r\lesssim r_N$ the logarithmic singularity is smeared out and the wave function remains smooth as shown in Fig.~\ref{fig3}(a). 

At any fixed $z$ singularity smearing occurs in the plane perpendicular to the $z$ axis at the distance $r\lesssim r_N$ from it. The region near 
the $z$ axis can be called {\it thread} of the state. The thread diameter is of the same order as the nuclear size. The resulting nuclear thread 
connecting two deuterons is sketched in Fig.~\ref{fig4}. 

Inside the thread the translational coordinate, with the typical energy $\hbar^2/(M_Dr^{2}_{N})\simeq 4.568~{\rm MeV}$, and nuclear internal 
degrees of freedom, with the typical energy $E_b$, are hybridized. These internal degrees are responsible for nucleons attraction but not for 
their inner structure. In other words, charge-mass distribution, associated with deuteron formation, is perturbed due to the hybridization with 
coordinates of center of mass. 
\subsection{Well at the thread}
The term (\ref{28BB}), being interpreted as $-(\hbar^2/M_D)\nabla^2$, corresponds to the peak of the wave function illustrated in 
Fig.~\ref{fig3}(a). Such peak should be compensated by a counter-term which is generic with a potential well and is associated with $\delta\psi$
in Eq.~(\ref{28A}).

It is impossible to describe this well solely in terms of the translational coordinate $\vec r$ since it is hybridized with the nuclear internal 
subsystem close to the thread. One can schematically draw this well as in Fig.~\ref{fig3}(b) where the zero energy level in Eq.~(\ref{28a}) is 
indicated by the thick line. At $r_N<r$ the plot goes over into $U$ in Eq.~(\ref{28a}). We keep this notation for all distances $r$. The well 
depth can be estimate as $E_b\sim\hbar^2/(M_Dr^{2}_{N})\sim 1~{\rm MeV}$. The underbarrier behavior, at $r_N<r$, is described in Sec.~\ref{form}.  

The given wave function (\ref{28aaa}) outside the thread results in $\alpha(\vec u_{\perp})$ in Eq.~(\ref{28A}) and the term (\ref{28BB}). 
The latter is the source for the part $\delta\psi$ (thread well). In other words, the driving force for the thread well is ultimately the given 
wave function (\ref{28aaa}). The wave function (\ref{28aaa}) is only compatible with the thread well. So there is a bridge between macroscopic 
and internal nuclear subsystems. This is unusual since the former are related to room temperature energy but the latter are associated with high 
energy of the nuclear scale. 
\subsection{Thread at short distances between deuterons}
Applicability of the quantum mechanical approach of Sec.~\ref{form} holds outside the thread, $r_N<r$. As follows from Eq.~(\ref{28a}), the 
typical $r$ is of the order of $R/B$. Therefore at $Br_N<R$ the logarithmic grow up of the wave function occurs outside the thread. At shorter 
distances between the deuterons, $R<Br_N$, there is only the exponential tail (\ref{15}) of the wave function outside the thread. This region is 
marked in Fig.~\ref{fig4}. 

In that case the potential $U$ in Eq.~(\ref{28a}) is increased resulting in that exponentially small underbarrier wave function. But the effective
zero energy level, indicated by the thick line in Fig.~\ref{fig4}, remains at the same position outside the thread. It is only possible if the 
potential well in Fig.~\ref{fig3}(b) is still deep that is of the order of 1~MeV. Therefore the wave function remains of the peak form as in
Fig.~\ref{fig3}(a) when deuterons are closer to each other, $R<Br_N$. So the thread state continues until two deuterons get in contact. 

At large $R$ the particle flux to the center is mainly determined by the region outside the thread, Sec.~\ref{form}(E). At $R<Br_N$ this flux 
is concentrated inside the thread. 
\subsection{Motion through the Coulomb barrier region}
The usual underbarrier penetration of a low-energy particle is due to its translational motion which is not bound to internal nuclear subsystem. 
This corresponds to WKB approach \cite{LANDAU1}.

In our case the mechanism is different. It is impossible to describe the phenomenon solely in terms of translational motion across the Coulomb 
barrier. The thread well is formed by the both translational and internal nuclear subsystems. One can say that the resulting potential is a 
combination of Coulomb and thread ones. The Coulomb barrier is dominated by the thread well and the motion toward the center occurs with no 
classical reflections. See Sec.~\ref{form}. 

In the translational motion along the thread two deuterons can get in contact despite they are of low energy. This is not an underbarrier 
motion but one along the region which became classically allowed due to the thread well.

Actually details of the wave function and the well close to the thread are not crucial and serve for interpretation of the mechanism. Namely how
the logarithmic singularity is cut off. The region outside the thread, described in Sec.~\ref{form}, allows to analyze the motion which is 
forbidden for low-energy WKB particles. It that region the wave function increases toward the thread which is equivalent to a negative 
transverse kinetic energy. This energy part dominates the Coulomb barrier as follows from Sec.~\ref{form}. The total particle flux to the center 
is mainly determined by the region outside the thread described in Sec.~\ref{form}.
\section{MACROSCOPIC NUCLEAR STATES}
\label{super} 
The angular part in Eq.~(\ref{28}) can be of a more general form. One can write $P_l(\vec n\vec n\,')$, 
where $\vec n\,'$ is an arbitrary unitary vector in the space. In this case the thread is localized along 
the $\vec n\,'$ direction. The total wave function outside the thread is of the type (\ref{28aaa})
\begin{equation}
\label{28b}
\psi(\vec R,\{\xi\},\vec n\,')=\frac{\Phi_l(R)}{R}P_l(\vec n\vec n\,')\psi_N\{\xi\},
\hspace{0.5cm}r_N<R\theta,
\end{equation}
where the Coulomb part is analogous to (\ref{28}) and $\{\xi\}$ is the symbolic notation for internal nuclear degrees of freedom. Outside the 
thread the translational motion and internal nuclear subsystem are separated. Inside the thread, $R\theta<r_N$, the separation does not hold and 
the total wave function $\psi(\vec R,\{\xi\},\vec n\,')$ is not of the form (\ref{28b}).

A continuous superposition of the above radial waves, propagating to the same point but with different 
directions in space, is also a state. It can be extended over a macroscopic space domain (say, of micron 
size). The wave function is mainly the product $\Psi_{C}(\vec R)\psi_{N}\{\xi\}$ originated from a spatial average 
of the regions outside threads. The difference of that expression from one, calculated for point charges \cite{LANDAU1}, is in the 
Coulomb part. $\Psi_{C}(\vec R)$ strongly differs from $\Psi^{(0)}_{C}(\vec R)$. 

That wave function does not obey the usual Schr\"{o}dinger equation in terms of translational coordinates. But there is no contradiction since 
the wave function is determined by the total system accounting for also internal nuclear subsystem. There is an additional contribution 
to the wave function where the translational motion and internal nuclear processes are not separated. This part originates from a spatial average 
of inner regions of threads. This contribution is small (as the nuclear radius) but it is connected to the high energy nuclear subsystem. As a 
result, the solution $\Psi^{(0)}_{C}(\vec R)$ of the usual Schr\"{o}dinger equation is substantially corrected.

So a motion of nuclei in a macroscopic region of space becomes completely different compared to the expectation based on the point charge approach.

One can introduce the complete state which is a superposition of ones with various thread directions in space. 
The complete wave function is
\begin{equation}
\label{28c}
\Psi(\vec R,\{\xi\})=\int f(\vec n\,')\psi(\vec R,\{\xi\},\vec n\,') do\,',
\end{equation}
where $f(\vec n\,')$ is the distribution of thread directions. Below we calculate the function (\ref{28c}) 
at $Br_N<R$. For this purpose one can use the expression (\ref{28b}) and to write
\begin{equation}
\label{29}
\Psi(\vec R,\{\xi\})=\frac{\Phi_l(R)}{R}\,Q(\vec n)\psi_N\{\xi\},\hspace{0.5cm}Br_N<R,
\end{equation}
where 
\begin{equation}
\label{30}
Q(\vec n)=\frac{B^2q^2}{2}\int f(\vec n\,')\,P_l(\vec n\vec n\,')\,do\,'.
\end{equation}
The integration in (\ref{30}) is extended outside the narrow cone along the direction $\vec n\,'$. The angle 
of this cone $\theta\sim r_N/R$ is much smaller than the typical scale of 
$f(\vec n\,')$ ($\delta\theta\sim 1$) and of $P_l(\vec n\vec n\,')$ ($\delta\theta\sim 1/B$). 

Suppose that the vector $\vec n\,'$ corresponds to the polar angle $\theta\,'$ and the azimuthal angle 
$\varphi\,'$. Then 
\begin{equation}
\label{31}
\vec n\vec n\,'=\cos\theta\cos\theta\,'+\sin\theta\sin\theta\,'\cos(\varphi\,'-\varphi)\,.
\end{equation}
Since the cone in Fig.~\ref{fig2} is narrow ($\theta\sim 1/B$), typical values of $(\theta-\theta\,')$ 
and $(\varphi-\varphi\,')$ are small and one can write
\begin{eqnarray}
\nonumber
&&Q(\theta,\varphi)=\frac{B^2q^2}{2}\int^{2\pi}_{0}d\varphi\,'\int^{\pi}_{0}d\theta\,'\sin\theta\,' 
f(\theta\,',\varphi\,')\\
\label{32}
&&\times P_l\left[1-\frac{(\theta-\theta\,')^2+(\varphi-\varphi\,')^2\sin^2\theta}{2}\right].
\end{eqnarray}
The function $f(\vec n\,')$ is smooth on the scale $1/B$. This allows to represent Eq.~(\ref{32}) in the 
form
\begin{eqnarray}
\nonumber
&&Q(\theta,\varphi)=\frac{B^2q^2}{2}\int^{\infty}_{-\infty}d(\varphi\,'
-\varphi)\sin\theta\int^{\infty}_{-\infty}d(\theta\,'-\theta)\\
\label{33}
&&\times P_l\left[1-\frac{(\theta-\theta\,')^2
+(\varphi-\varphi\,')^2\sin^2\theta}{2}\right]f(\theta,\varphi)\,,
\end{eqnarray}
which is
\begin{equation}
\label{34}
Q(\theta,\varphi)=\pi B^2q^2f(\theta,\varphi)\int^{\infty}_{\theta_0}P_l(\cos{\tilde\theta})\,
{\tilde\theta}\,d{\tilde\theta}\,.
\end{equation}
Here ${\tilde\theta}^{2}=(\theta\,'-\theta)^2+(\varphi\,'-\varphi)^2\sin^2\theta$ and 
$\theta_0\sim r_N/R$. ${\tilde\theta}$ is a small angle between vectors $\vec n\,'$ and $\vec n$. The 
integration leads to $Q\simeq f(\theta,\varphi)$, where corrections to this relation are proportional to 
$r_N$. 

The complete wave function can be written in the form
\begin{equation}
\label{35}
\Psi(\vec R,\{\xi\})=\Psi_C(\vec R)\psi_N\{\xi\}+\Psi_{CN}(\vec R,\{\xi\}), 
\end{equation}
where the Coulomb part is
\begin{equation}
\label{36}
\Psi_C(\vec R)=\frac{\Phi_l(R)}{R}f(\vec n)    
\end{equation}
and $\vec R=R\vec n$. The small part $\Psi_{CN}(\vec R,\{\xi\})$ is added which is absent in Eq.~(\ref{28c}).
This part comes from thread regions where macroscopic and internal nuclear subsystems are hybridized.

The first part in the right hand side of (\ref{35}) exists at $Br_N\ll R$ due to 
limits of integration in Eq.~(\ref{34}). This part comes from the integration outside the thread. The additional (translational-nuclear) 
part $\Psi_{CN}$ comes from the thread region, ${\tilde\theta}<r_N/R$, where macroscopic and nuclear subsystems are not separated.

Translational and internal nuclear subsystems are hybridized in space providing {\it macroscopic nuclear state}. For the isotropic case 
(a constant $f(\vec n)$) the wave function (\ref{36}) satisfies the equation in physical units
\begin{equation}
\label{37}
-\frac{\hbar^2}{M_D}\frac{1}{R^2}\frac{\partial}{\partial R}\left(R^2\frac{\partial\Psi_{C}}{\partial R}\right)
+\left[\frac{e^2}{R}+U_{CN}(R)\right]\Psi_C=E\Psi_C
\end{equation}
where the ``ghost'' potential is $U_{CN}=-q^2e^4/(ER^2)$. It comes from a rearrangement in the nuclear subsystem but exists on a macroscopic scale. 
\begin{figure}
\includegraphics[width=6cm]{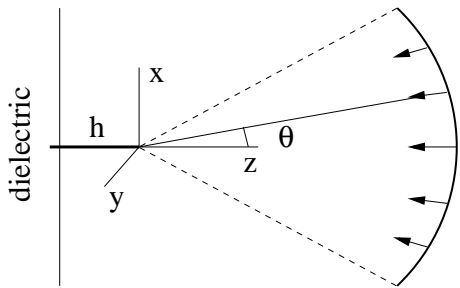}
\caption{\label{fig5}Scheme for creation of the nuclear thread at $z<0$ shown by the thick line. The spherical incident flux at large distances 
should be of the certain form as a function of the angle $\theta$.}
\end{figure}
\section{A WAY TO CREATE THE NUCLEAR THREAD}
\label{exp}
In previous sections the infinitely long thread is studied. But in a usual flux of deuterons translational and internal nuclear motions are not 
hybridized since the usual wave function (plane wave, for example) does not have a tendency for singularity on some line. Suppose that at large 
distances there is no hybridization but the incident flux is more complicated than, say, a plane wave. One can put a question: Is it possible to 
get such flux hybridized sufficiently close to a target? We show in this section that the answer is positive and we specify the forms of such 
fluxes.

We consider not counter-beams of deuterons but the interaction of them with the plane dielectric wall shown in Fig.~\ref{fig5}. The related mirror 
force interaction \cite{LANDAU2} is
\begin{equation}
\label{38}
V(r,z)=-\frac{(\varepsilon-1)e^2}{4(\varepsilon+1)(h+z)}\,,\hspace{0.5cm}\lambda\ll(h+z),
\end{equation}
where $r^2=x^2+y^2$ and $\varepsilon$ is the dielectric constant. We put the zero coordinate on the distance $h$ from the dielectric. One can 
choose $h\sim e^2/E\sim 500\AA$ according to the Coulomb scale of the problem. The interaction (\ref{38}) holds until $(h+z)$ exceeds the atomic 
size which is of the order of $\lambda\sim 1\AA$. Closer to the dielectric $V$ becomes $r$ dependent.
\subsection{Equations}
The Schr\"{o}dinger equation for the particle with the mass $M_D$ in the potential (\ref{38}) is
\begin{equation}
\label{39}
-\frac{\partial^2\psi_{C}}{\partial z^2} -\frac{1}{r}\frac{\partial}{\partial r}\left(r\frac{\partial\psi_{C}}{\partial r}\right)
-\frac{\alpha}{h+z}\psi_C=\psi_C\,,
\end{equation}
where 
\begin{equation}
\label{40}
\alpha=\frac{B(\varepsilon-1)}{2(\varepsilon+1)\sqrt{2}}
\end{equation}
and distances are measured in the units of the wave length $\lambda=\hbar/\sqrt{2EM_D}$. So $\alpha\sim h\sim B$. 
\begin{figure}
\includegraphics[width=6cm]{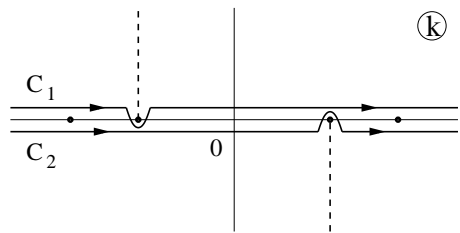}
\caption{\label{fig6} Contours of integration in the complex $k$ plane. The cuts (dashed lines) are terminated at branch points at $k=\pm 1$. The 
poles at $k=\pm\sqrt{1+q^2}$ are marked by points.}
\end{figure}

Eq.~(\ref{39}) allows separation of variables and the solution is of the type
\begin{equation}
\label{41}
f(z,k)H^{(2)}_{0}\left(-ir\sqrt{k^2-1}\right),
\end{equation}
where $H^{(2)}_{0}$ is the Hankel function \cite{GRAD}. The function $f(z,k)$ satisfies the equation
\begin{equation}
\label{42}
-\frac{\partial^2 f}{\partial z^2}-\frac{\alpha}{z+h}f=k^2 f.
\end{equation}
A solution of Eq.~(\ref{42}) can be written as
\begin{equation}
\label{43}
f(z,k)=f_{1}(z,k)+f_{2}(z,k),
\end{equation}
where, at real large $k$ of any sign, $f_1(z,k)\sim\exp(ikz)$ and $f_2(z,k)\sim\exp[-ik(z+2h)]$. The wave function is $\psi_C(r,z)=\psi_{C1}(r,z)
+\psi_{C2}(r,z)$, where
\begin{equation}
\label{44}
\psi_{C1,2}(r,z)=\int_{C_{1,2}}dkA(k)f_{1,2}(z,k)H^{(2)}_{0}\left(-ir\sqrt{k^2-1}\right).
\end{equation}
Integration contours in the complex $k$ plane are shown in Fig.~\ref{fig6}. The contours bypass the branch points of the function $\sqrt{k^{2}-1}$ 
and the cuts are drawn by dashed lines. The function $\sqrt{k^{2}-1}$ is real and positive at large positive $k$. 

The weight function $A(k)$ is of the form
\begin{equation}
\label{45}
A(k)=\frac{\delta}{\sinh\left[(k+\sqrt{1+q^2})\delta\right]}+\frac{\delta}{\sinh\left[(k-\sqrt{1+q^2})\delta\right]}
\end{equation}
where the real $q\sim 1$ is chosen. The parameter $q$ is analogous to one in Sec.~\ref{form}(A) and it is also a characteristic number of the 
problem determined by the incident flux. $q$ controls decay of the wave function away from the singularity line. The meaning of the parameter 
$\delta$ is clarified below.
\subsection{Solution}
On the basis of above equations one can analyze the solution at various distances. 
\subsubsection{Wave function at large distances}
At real $k$ and large $z$ $f_1(z,k)=\exp[ikz+i\chi_1(z,k)]$, where $\chi_1(z,k)$ is the certain phase depending on $z$ as $(\alpha/2k)\ln z$ (the 
Coulomb phase) \cite{LANDAU1}. With the use of the asymptotics of the Hankel function \cite{GRAD} the rapidly oscillating integrand in (\ref{44}) 
for $\psi_{C1}$ is
\begin{equation}
\label{48}
\exp\left(izk-r\sqrt{1-k^2}\right)\simeq\exp\left[-iR+\frac{iR}{2\sin^2\theta}(k-k_0)^2\right]
\end{equation}
where $R^2=z^2+r^2$ and the saddle point is
\begin{equation}
\label{49}
k_0(r,z)=-\frac{z}{\sqrt{z^2+r^2}}=-\cos\theta.
\end{equation}
The polar angle $\theta$ is shown in Fig.~\ref{fig5}. Making the steepest descent integration in Eq.~(\ref{44}) we obtain at large $R$ and 
positive $z$ ($0<\theta<\pi/2$)
\begin{equation}
\label{50}
\psi_{C1}=\frac{2i}{R}\,A(\cos\theta)\exp\left[-iR+i\chi_1(z,-\cos\theta)\right]
\end{equation}
Analogous simple calculations can be done for $\psi_{C2}$ in Eq.~(\ref{44}) where $f_2(z,k)=\exp[-ik(z+2h)+i\chi_2(z,k)]$ and $\chi_2(z,k)$ depends 
on $z$ as $-(\alpha/2k)\ln z$ at large $z$. The saddle point is $\cos\theta$ resulting in
\begin{equation}
\label{51}
\psi_{C2}=\exp\left[2i\varphi(\cos\theta)\right]\psi_{C1},
\end{equation}
where $\varphi(\cos\theta)$ depends on $\chi_1(z,-\cos\theta)-\chi_2(z,\cos\theta)$ and therefore it is $z$ independent.
\subsubsection{Singularity formation}
The wave function can be written in the form
\begin{equation}
\label{52}
\psi_{C}(r,z)\simeq\Phi(z)\ln r+({\rm non\,\,singular\,\,part}).
\end{equation}
The $\ln r$ singularity is smeared out on the nuclear size. This mechanism is described in Sec.~\ref{cone}. Between the dielectric and the 
point $z=0$ the singularity strength $\Phi(z)$ is not small. This function exponentially decays at positive $z$. The goal is to investigate 
formation of such type of singularity and to conclude which particle flux has to be at large positive $z$ to realize that scenario. 

The singularity strength in Eq.~(\ref{52}) is given by the relation
\begin{equation}
\label{53}
\Phi(z)=-\frac{2i}{\pi}\left[\int_{C_{1}}dkA(k)f_1(z,k)+\int_{C_{2}}dkA(k)f_2(z,k)\right].
\end{equation}
Since $A(k)$ (\ref{45}) exponentially decreases at large wave numbers $k$, $\Phi(z)$ is smooth even close to the point $z=0$. 

It is instructive to check Eq.~(\ref{53}) for a free particle ($\alpha=0$). In this case $f_1(k)=\exp(ikz)$, $f_2(k)=0$, and we obtain
\begin{equation}
\label{54}
\Phi(z)=2i\pi\cos(z\sqrt{1+q^2})\left(\tanh\frac{\pi z}{2\delta}-1\right),\hspace{0.5cm}\alpha=0.
\end{equation}
The meaning of the parameter $\delta$ is clear from Eq.~(\ref{54}). It is the width of the region on the $z$ axis where $\ln r$ singularity is 
formed. At small $\delta$ the function $\Phi(z)$ in (\ref{54}) is proportional to $\theta(-z)$ and Eq.~(\ref{52}) is the part of the form 
$\psi_C(r,z)=-2i\pi\ln(\sqrt{z^2+r^2}+z)$. As one can show, this formula follows from bypasses of the cuts in Fig.~\ref{fig6} after contours 
deformation.

One can treat $f_1(z,k)$ as some incident wave and $f_2(z,k)$ as a reflected one. $f_{1,2}(z,k)$ are regular functions of the variable $k$ in the 
entire complex $k$ plane excepting poles of $f_{2}(z,k)$ \cite{LANDAU1,REGGE}. These poles are analogous to Regge ones and correspond to large 
${\rm Im}\,k\sim\alpha$. Bypasses of them relate to the exponentially small contribution $\exp(-{\rm const}\,\alpha)$. The integrand in (\ref{44}) 
for $\psi_{C1}$ exponentially decays in the upper half-plane of $k$ when $z$ is positive. At negative $z$ it decays in the lower half-plane. The 
integrand for $\psi_{C2}$ decays in the lower half-plane for all $z$.

In our case at positive $z\gg\delta$ the main contribution to $\Phi(z)$ is given by the first term in (\ref{53}) where one should account for the 
nearest complex pole in the upper half-plane of $k$. This gives $\Phi(z)\sim\exp(-\pi z/\delta)$ which is similar to (\ref{54}).

At $-h<z<0$ the singular part, determined by the poles $k=\pm\sqrt{1+q^2}$ associated with $\psi_{C1}$, is
\begin{equation}
\label{55}
\psi_C(r,z)=4K_0(qr)f_1(z,\sqrt{1+q^2}),
\end{equation}
where $K_0(qr)=-(i\pi/2)H^{(2)}_{0}(-iqr)$ \cite{GRAD} and 
\begin{equation}
\label{56}
K_0(qr)=
\begin{cases}
-\ln(qr/2)-0.577,\hspace{0.5cm}qr\ll 1\\
\sqrt{\pi/2qr}\exp(-qr),\hspace{0.5cm}1\ll qr.
\end{cases}
\end{equation}
In Eq.~(\ref{55}) the modulus of $f_1$ is of the order of unity.

Not too close to the dielectric surface, $\lambda<(h+z)$, the deuteron is attracted to it according to the potential (\ref{38}). At shorter 
distances, which are less than the atomic size, the Coulomb repulsion is dominated by thread effects as at $z<Br_N$ in Fig.~\ref{fig4}. 
Analogously to that, the thread, created outside the dielectric, continues inside it.
\subsubsection{Complete wave function}
The non-singular part of the wave function (in addition to (\ref{55})) is complex and therefore the current 
$2\pi r\,{\rm Im}(\psi^{*}_{C}\partial\psi_{C}/\partial r)$ toward the thread is finite at $r\rightarrow 0$. This effect results in an nonphysical 
increase of the flow along the core. Also, as follows from Eqs.~(\ref{50}) and (\ref{51}), at large distances there is solely the incident wave 
and there is no reflected one. To provide the zero current toward the thread and the reflected wave we take ${\rm Re}\,\psi_C$ as a wave function. 
In the case of a free particle in two dimensions this is similar to the choice of $N_{0}(kr)$ which is logarithmically singular at small $r$ 
\cite{GRAD}. Not that the choice of ${\rm Re}\,i\psi_C$ would correspond to $J_0(kr)$ which is not our case.
\subsubsection{Incident flux at large distances}
Here we use physical units by substitution of $R$ in previous equations by $R/\lambda$. The incident radial flux density is \cite{LANDAU1}
\begin{equation}
\label{57}
j_R=\frac{\hbar}{M_D}{\rm Im}\left(\psi^{*}_{C}\frac{\partial\psi_C}{\partial R}\right),
\end{equation}
where $\psi_C$ is the sum of functions (\ref{50}) and (\ref{51}). The total incident flux $I_R$ satisfies the relation
\begin{equation}
\label{58}
\frac{dI_R}{d\theta}=2\pi R^2\sin\theta\,j_R.
\end{equation}
Let us multiply the wave functions (\ref{50}) and (\ref{51}) by a real constant $a$. Then one can easily calculate from Eqs.~(\ref{57}), (\ref{58}), 
(\ref{50}), and (\ref{51}) that 
\begin{equation}
\label{59}
\frac{dI_R}{d\theta}=-I_0 A^2(\cos\theta)\sin\theta\cos^{2}\left[\varphi(\cos\theta)\right].
\end{equation}
The parameter $I_0$, which has the dimensionality of inverse time, is connected to $a$ by the relation
\begin{equation}
\label{60}
a=\frac{1}{8\lambda^{3/2}}\sqrt{\frac{\hbar I_0}{\pi E}}\,.
\end{equation}
If to choose the parameter $\delta$ to be small the angular distribution of the incident flux takes the form
\begin{equation}
\label{61}
\frac{dI_R}{d\theta}=-I_0 \,\frac{4\sin\theta\cos^2\theta}{(q^2+\sin^2\theta)^2}\cos^{2}\left[\varphi(\cos\theta)\right].
\end{equation}

The angular dependence of the incident flux (\ref{61}) is a product of the smooth envelope function and the rapidly oscillating part 
($\cos^2\varphi$) leading to fringes with small $\Delta\theta\sim \lambda/h$. These fringes appear automatically due to the interference with 
the reflected wave passing the distance $2h$. 

The wave function should be multiplied by the factor (\ref{60}). On the basis of that one can estimate the number of deuterons penetrating inside 
the dielectric within the thread of the radius $r_N$ and the length $\lambda$
\begin{equation}
\label{62}
N=\lambda r^{2}_{N}|\psi_{C}|^2\sim\frac{\hbar}{E}\left(\frac{r^{2}_{N}}{\lambda^2}I_0\right),
\end{equation}
where $\psi_C$ is given by Eq.~(\ref{55}) and $I_0\,r^{2}_{N}/\lambda^2$ is the fraction of the total flux $I_0$ propagating inside the thread. 
There is the equal counter-flow (Sec.~\ref{exp} B, part 3) and the state remains steady. 

That description holds when deuterons do not go out of the game due to possible nuclear reactions at short distances. If these reactions would 
occur they reduce the deuterons number with the rate $N/t$. $t$ is the nuclear time when $\hbar/t$ is of the order of the Coulomb energy $e^2/r_N$. 
So the flow of deuterons, which may disappear due to possible reactions, is 
\begin{equation}
\label{63}
I\sim\frac{r_N}{r_B}I_0.
\end{equation}
Now the incident flux $I_0$ exceeds the reflected one by the part (\ref{63}) and the total wave function is not exactly $\psi_C+\psi^{*}_{C}$. 
This is equivalent to a finite flux to the thread pointed out in Sec.~\ref{exp} B, part 3. So the extra flux of deuterons (\ref{63}) does not enter 
inside the thread through its cross-section of the area $\pi r^{2}_{N}$ but through the side surface of the thread. 
\subsection{How to create the nuclear thread}
We conclude that nuclear hybridization close to the target is determined by properties of the flux at large distances where there is no 
hybridization (the negligible exponential tail). The transition to well developed (at $z<0$) hybridization occurs in a soft manner within the 
interval $\delta$ around the point $z=0$. Therefore the nuclear subsystem enters the game smoothly with distance. 

A weak deviation of the envelope from the optimal form (\ref{61}) does not influence formation of the thread. The form (\ref{61}) is not a unique 
one for thread formation. In this paper we do not study other possibilities.

In experiments one should be two conditions. (i) {\it The incident flux has to be of the radial form at large distances directed to some point in 
front of the dielectric surface}. This is shown in Fig.~\ref{fig5}. (ii) {\it The angular dependence of the radial flux should be close to the 
optimal envelope} (\ref{61}).

Possible components with $m\neq 0$ (not axially symmetric states) correspond to the usual parts $r^m\exp(im\varphi)$ of the wave function and do 
not influence the scenario.

It is not necessary to keep the particle flux (\ref{61}) for all angles. It can be of that form solely within some part of the total sphere as 
shown in Fig.~\ref{fig5}. From this part the wave function continues toward the point $R=0$ providing the singularity formation. The component of 
the total wave function, associated with the border of this spherical part, is not singular at $R=0$. 

It follows that manipulation with an incident flux of low energy deuterons can lead to their reactions with a not small probability. We study the 
incident flux which results just in the single thread. One can also adjust the flux to get a set of threads perpendicular to the dielectric. 
\section{DISCUSSIONS}
\label{disc}
Usually the wave function of the translational motion is smooth on the nuclear scale and the nucleus can be considered as a point charge. In this 
case the translational motions is well separated from internal nuclear subsystem. 

It happens that the opposite limit is possible. For the motion in the Coulomb field the wave function of the translational motion increases 
toward the line connecting the nucleus and another Coulomb center. This wave function is singular if to treat the nucleus as a point charge. 

In reality, for finite $r_N$, the Coulomb interaction leads to non-separation of translation and internal nuclear coordinates. The latter 
provide spatial fluctuations in the singular state resulting in its smearing. This is analogous to the electron dynamics in an external 
field, mediated by photons, in quantum electrodynamics. In the both cases the particle momentum is not conserved in virtual processes leading 
to spatial ``oscillations'' of the center of mass.

Due to the distribution of singularity lines the wave function of the deuteron translational motion becomes smooth and having a 
peak. The peak of the wave function is localized within the cylinder of the nuclear radius called thread. This localization is an indication of 
the certain potential well in the thread. Such a well is formed with the participation of nuclear internal subsystem related to the energy of 
the order of $E_b$. The driving force for the well is a special form of the incident particle flux at large distances.

So in this manner the bridge is built up between the macroscopic motion and internal nuclear subsystem. This is unusual since the former are 
related to room temperature energy but the latter are associated with high energy of the nuclear scale. 

The conventional description solely in terms of translational motion across the Coulomb barrier is impossible. The resulting potential is a 
combination of Coulomb and thread ones. The Coulomb barrier is dominated by the thread well and the motion toward the center occurs with no 
classical reflections. The wave function increases approaching the thread which is equivalent to the negative transverse kinetic energy. This 
energy dominates the Coulomb barrier and the nucleus moves along the thread with no reflections. Nuclear thread can be created at some region 
by adjusting a deuteron flux at large distances.

In the translational motion along the thread two deuterons can get in contact despite they are of low energy. This is not an underbarrier 
motion but one along the region which became classically allowed due to the thread well. 

A superposition of nuclear states with various directions of threads is also a state of the system of two deuterons. It can be macroscopically 
extended in the space (macroscopic nuclear state). The wave function of macroscopic nuclear state is mainly a product of the Coulomb part, 
$\Psi_C(\vec R)$, and the nuclear one, $\psi_N\{\xi\}$. In addition, there is a small (proportional to the deuteron size) Coulomb-nuclear 
part $\Psi_{CN}(\vec R,\{\xi\})$ where the translational motion and nuclear internal subsystem are not separated. 

When an inter-nuclear distance is macroscopically larger than the nuclear size, there is a normal way to 
describe nuclei in such three-dimensional macroscopic region. One should write down the usual 
Schr\"{o}dinger equation for point charges interacting via their Coulomb field. The result for two nuclei 
is well described in textbooks. 

It is surprising to say that these famous results sometimes can be dramatically incorrect and we encounter
the paradoxical phenomenon. In a nuclear macroscopic state the spatial distribution of nuclei is against to 
what is expected from the usual Schr\"{o}dinger equation based on the point charge approach. Nuclei "do not 
satisfy`` Schr\"{o}dinger equation. It is not a contradiction since the wave function is determined by the total system accounting 
for also internal nuclear processes.

One of the ways to form the hybridized state is to create the radial flux (with the certain angular distribution) of deuterons directed at
some point close to the dielectric surface. We will discuss elsewhere particular ways of experimental formation of proper fluxes. One can just 
list experimental possibilities for this. {\it Confinement beams} arrangements. The use of {\it quantum lens}. One can push deuterons to pass 
through a diffraction grid of a shape to be adjusted. The use of {\it magnetic} and {\it electric} fields for formation of fluxes. One can use a 
{\it not hot plasma}. In this case nucleus wave function is a superposition of various states including thread like ones.
\section{CONCLUSIONS}
At large distances one can adjust the incident flux to get the thread state at a shorter distance between nuclei. The tread well is formed by the 
both translational and internal nuclear subsystems. The Coulomb barrier is dominated by the thread well. 

In this manner the bridge is built up between macroscopic and internal nuclear subsystems. This is unusual since the former are related to room 
temperature energy but the latter are associated with high energy of the nuclear scale. 

In the translational motion along the thread two deuterons can get in contact despite they are of low energy. This is not an underbarrier 
motion but one along the region which became classically allowed due to the thread well. 

\acknowledgments
I appreciate valuable discussions with S. J. Brodsky and H. C. Rosu.
\appendix*
\section{ELECTRON THREADS}
Instead of two deuterons one can consider two electrons and try to construct the thread state of them. In this section we mention some aspects 
of electron thread formation in quantum electrodynamics.
\subsection{General arguments}
An electron, acted by an external macroscopic field, is spatially smeared due to the interaction with zero point electromagnetic oscillations. 
If $\vec u$ is the displacement of the electron center of mass from its mean field position then $\langle\vec u\rangle=0$ and 
$\langle u^2\rangle$ is finite. 

In an external field (for example Coulomb one) due to the uncertainty in electron positions it probes various parts of the potential. This 
changes its energy. The famous example of this phenomenon is the Lamb shift of atomic levels \cite{LANDAU3}. 

Conditions for formation of nuclear thread states, discussed in the paper, can be used in the analogous problem of creation of electron threads.
For thread formation the main point is cutting off the logarithmic singularity described in terms of coordinates of the electron center of mass. 
The singularity is smeared out by fluctuations of those coordinates $\sqrt{\langle u^2\rangle}$. For the electron thread that distance plays a 
role of $r_N$. 
\subsection{Estimate of the mean squared displacement of the electron}
Unlike the Lamb shift, it is impossible to apply the perturbation theory to calculate smearing of a singular wave function of electron. In that 
problem one should calculate all orders of the perturbation theory. For this reason, we use below the approximate method just to estimate the
electron mean squared displacement \cite{MIGDAL}. The method is successfully applied for study of the Lamb shift. In this method the electron 
motion under the action of zero point oscillations can be described by the equation
\begin{equation}
\label{A1}
m\frac{d^2\vec u}{dt^2}+m\Omega^2\vec u=-e\vec{\cal E},
\end{equation}
where $m$ is the electron mass and $\Omega\sim me^4/\hbar^3$ is the electron rotation frequency in the atom. 

One can use the Fourier expansion 
\begin{equation}
\label{A2}
\vec u(\vec R,t)=\sum_{k}\vec u_{k}\exp(i\vec k\vec R-i\omega_{k}t)
\end{equation}
and analogous one for the fluctuating electric field $\vec{\cal E}(\vec R,t)$. Since $\vec u(\vec R,t)$ is real it should be 
$\vec u^{*}_{k}=\vec u_{-k}$ and $\omega_{-k}=-\omega_{k}$ in the expansion (\ref{A2}). The condition $uk\ll 1$ has to be held in this method.
The solution of Eq.~(\ref{A1}) is of the form
\begin{equation}
\label{A3}
\vec u_{k}=\frac{e\vec{\cal E}_{k}}{2m|\omega_k|}\left(\frac{1}{|\omega_k|+\Omega}+\frac{1}{|\omega_k|-\Omega}\right).
\end{equation}
According to the quantum mechanical approach, Eq.~(\ref{A3}) should be modified as
\begin{equation}
\label{A4}
\vec u_{k}=\frac{e\vec{\cal E}_{k}}{2m|\omega_k|}\left(\frac{\sqrt{1+n_k}}{|\omega_k|+\Omega}+\frac{\sqrt{n_k}}{|\omega_k|-\Omega}\right),
\end{equation}
where $n_k$ is the number of quanta, the first term relates to the quanta emission, and the second one to the absorption.

The mean squared displacement is 
\begin{equation}
\label{A5}
\langle u^2\rangle=\int\frac{d^{3}R}{V}\langle u^2\rangle=\sum_{k}\langle|\vec u_{k}|^2\rangle,
\end{equation}
where $V$ is the volume of the system. Since in our case $n_k=0$, the mean squared displacement has the form 
\begin{equation}
\label{A6}
\langle u^2\rangle=\frac{e^2}{4m^2}\sum_{k}\frac{\langle|\vec{\cal E}_k|^2\rangle}{\omega^{2}_{k}(|\omega_k|+\Omega)^2}\,.
\end{equation}
The energy of zero point oscillations is
\begin{equation}
\label{A7}
\int\frac{d^3R}{4\pi}\langle{\cal E}^2\rangle=\frac{V}{4\pi}\sum_{k}\langle|\vec{\cal E}_k|^2\rangle=\sum_{k}\frac{\hbar|\omega_k|}{2}.
\end{equation}
It follows from here that $\langle|\vec{\cal E}_k|^2\rangle=2\pi\hbar|\omega_k|/V$. Using the summation rule
\begin{equation}
\label{A9}
\sum_{k}=2\int\frac{4\pi k^2dkV}{(2\pi)^3}
\end{equation}
(the coefficient 2 accounts for two photon polarizations) and the relation $\omega_k=ck$, one can obtain from Eq.~(\ref{A6}) \cite{MIGDAL}
\begin{equation}
\label{A10}
\langle u^2\rangle=\frac{r^{2}_{B}}{\pi}\left(\frac{e^2}{\hbar c}\right)^3\int^{\omega_{max}}_{0}\frac{\omega d\omega}{(\omega+\Omega)^2}\,,
\end{equation}
where $r_B=\hbar^2/(me^2)$ is the electron Bohr radius. The upper limit $\omega_{max}$ is determined by the condition of non-relativistic motion,
that is $\omega_{max}\simeq mc^2/\hbar$. In the relativistic region $\vec u_k$ decreases due to the enhancement of the relativistic mass. Finally
\begin{equation}
\label{A11}
\langle u^2\rangle=\frac{r^{2}_{B}}{\pi}\left(\frac{e^2}{\hbar c}\right)^3\ln\frac{mc^2}{\hbar\Omega}\,.
\end{equation}
If to consider not an atom but another field the cut off $\hbar\Omega$ will be determined by this field. The exact cut off is not crucial for 
our estimate. The applicability condition of the approach used $\langle u^{\,2}\rangle k^{2}_{max}\sim\langle u^{\,2}\rangle/r^{2}_{c}\ll 1$ is 
valid since $\langle u^{\,2}\rangle/r^{2}_{c}\sim e^2/(\hbar c)$. Here $r_c=\hbar/(mc)\simeq 3.86\times 10^{-11}{\rm cm}$ is the electron Compton 
length. 

Note that the Coulomb potential of the atom, $-e^2/|\vec R+\vec u|$, being expanded up to the second order in $u$, produces the Lamb shift 
\cite{MIGDAL} coincided (excepting the numerical coefficient) with the exact result \cite{LANDAU3}. 

The distance $\sqrt{\langle u^2\rangle}\sim 10^{-11}{\rm cm}$ in quantum electrodynamics is close to the electron Compton length. One can put a 
general question on formation of the electron thread under certain conditions. The diameter of this thread is supposed to be $10^{-11}{\rm cm}$. 
It may be connected with two electrons or with one in some field, for example, in molecules and solids. In this case a specificity of associated 
chemical bonds are worth to be analyzed.

\end{document}